\documentclass{PoS}
\newcommand{\Olejnik}{{\v{S}}.~Olejn{\'\i}k}
\title{Gluon chains and the quark-antiquark potential%
\addtocounter{footnote}{1}\thanks{This research was supported in part by the U.S.\ DOE under Grant No.\ DE-FG03-92ER40711 (J.G.), and by the Slovak Grant Agency for Science, Project VEGA No.\ 2/0070/09, by ERDF OP R\&D, Project CE QUTE ITMS~26240120009, and via CE SAS QUTE (\v{S}.O.)}}

\ShortTitle{Gluon chains and the quark-antiquark potential}

\author{Jeff Greensite\\
Physics and Astronomy Dept., San Francisco State University, San Francisco, CA 94132, USA\\
E-mail: \email{jgreensite@gmail.com}}

\author{\addtocounter{footnote}{-2}\speaker{\v{S}tefan Olejn{\'\i}k}\\
        Institute of Physics, Slovak Academy of Sciences, SK--845 11 Bratislava, Slovakia\\
        E-mail: \email{stefan.olejnik@savba.sk}}

\abstract{The flux tube between a quark and an antiquark in Coulomb gauge is
imagined in the gluon-chain model as a sequence of constituent gluons
bound together by Coulombic nearest-neighbor interactions. We
diagonalize the transfer matrix in SU(2) lattice gauge theory in a
finite basis of states containing a static quark-antiquark pair together
with zero, one, and two gluons in Coulomb gauge.  We show that while the
string tension of the color-Coulomb potential (obtained from the
zero-gluon to zero-gluon element of the transfer matrix) overshoots the
true asymptotic string tension by a factor of about three, the inclusion
of a few states with constituent gluons reduces the discrepancy
considerably. The minimal energy eigenstate of the transfer matrix in
the zero-, one-, and two-gluon basis exhibits a linearly rising
potential with the string tension only about 1.4 times larger than the
asymptotic one.}

\FullConference{The XXVII International Symposium on Lattice Field Theory - LAT2009\\
		 July 26--31, 2009\\
		 Peking University, Beijing, China}

\begin{document}

\section{Color-Coulomb potential and gluon-chain model}\label{intro}
The color-Coulomb potential, i.e.\ the $R$-dependent part
of the energy of a physical state of a heavy quark-antiquark pair 
at distance $R$ in Coulomb gauge, was shown to represent an upper bound on the true static quark-antiquark potential \cite{Zwanziger:2002sh}.
Numerical simulations demonstrated that it was asymptotically linear and its string tension, $\sigma_{\mathrm{Coul}}$, was measured to be 2--3 times larger 
than the standard asymptotic string tension, $\sigma_{\mathrm{asymp}}$ 
\cite{Greensite:2003xf,Nakagawa:2006fk}.

In this context, a series of questions naturally arises: How do flux tubes form in the Coulomb gauge? 
What mechanism is behind the collimation of color-electric 
fields into flux tubes? How does it reduce $\sigma_{\mathrm{Coul}}$ to $\sigma_{\mathrm{asymp}}$?

\begin{figure}[b!]
\centerline{\includegraphics[width=\textwidth]{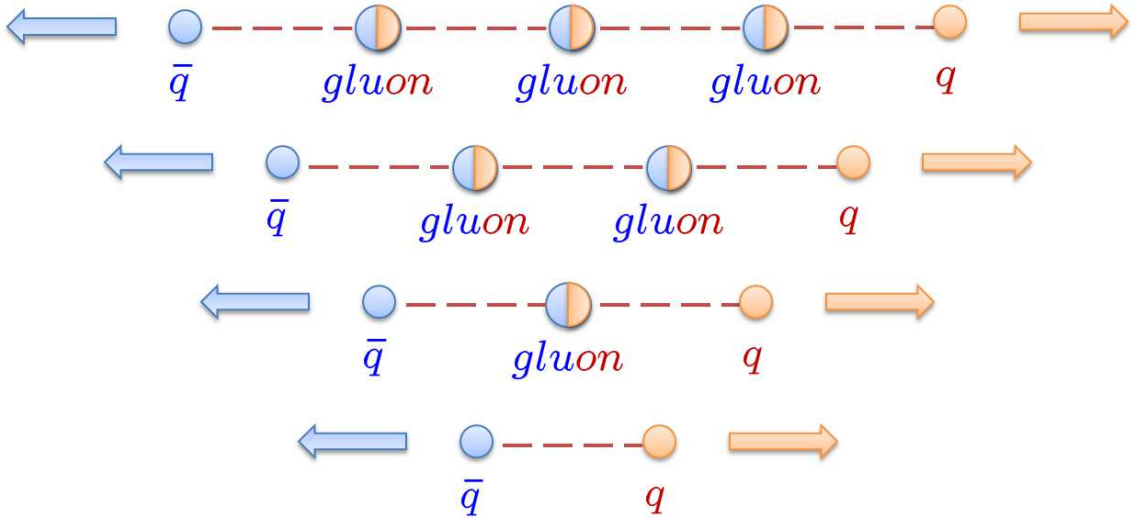}}
\caption{The gluon-chain model.}\label{chain-model}
\end{figure}

In the gluon-chain model, proposed first by Tiktopoulos \cite{Tiktopoulos:1976sj} and later developed by Thorn and one of the present authors (J.G.) \cite{Greensite:2001nx}, the flux tube between a quark and an antiquark is visualized as a sequence of gluons. As a heavy quark and antiquark move 
away from each other, a chain of constituent gluons arises between them. The constituent gluons are bound together by Coulombic nearest-neighbor 
interactions. Schematically (see Fig.\ \ref{chain-model}):
\begin{equation}
{\vert\Psi_{q\bar{q}}^\mathrm{chain}\rangle=
\bar{q}(0)\;\Big\{\alpha_0+\alpha_1 A+
\alpha_2 A A+\alpha_3 A A A+\dots\Big\}\;q(R)\;
\vert\Psi_{0}\rangle}.
\end{equation}

The goal of the present study was to test, in first-principle numerical lattice
Monte Carlo simulations, the conjecture that constituent gluons do reduce the magnitude of the static quark-antiquark
potential, and to measure the constituent gluon content of the QCD flux tube 
in Coulomb gauge. (For an early attempt in the same direction see Ref.\ \cite{Greensite:1988tj}.) We will mainly report
results of simulations of SU(2) lattice gauge theory at $\beta=2.4$ on a $22^4$ 
lattice. A complete set of our data is contained in Ref.\ \cite{Greensite:2009mi}.

\section{Transfer matrix and static potential}\label{outline}
The (Euclidean) time evolution in lattice gauge theory in Coulomb (or some other physical) gauge is governed by the transfer matrix
\begin{equation}
\mathcal{T}=\exp[-\mathcal{H}a], 
\end{equation}
where $\mathcal{H}$ is the Hamiltonian in that gauge, and $a$ the lattice spacing. We used its rescaled form:
\begin{equation}
\mathfrak{T}=\frac{\mathcal{T}}{\langle \Psi_0\vert\mathcal{T}\vert\Psi_0\rangle}=\exp[-(\mathcal{H}-E_0)a],
\end{equation}
where $\Psi_0$ denotes the ground state and $E_0$ is its energy.

In an ideal case one would like to diagonalize $\mathfrak{T}$ in the (infinite-dimensional) subspace of states containing a static 
quark-antiquark pair separated by a distance $R$. Then, the static quark-antiquark potential (in lattice units) could be computed from
\begin{equation}
V(R)=-\log(\tau_{\mathrm{max}}),
\end{equation}
where $\tau_{\mathrm{max}}$ is the largest eigenvalue of the rescaled transfer matrix $\mathfrak{T}$ in this subspace, corresponding to the minimal-energy eigenstate.
In reality, this program cannot be realized; instead one has to reduce the subspace of states considered to a manageable size. Fortunately, if the quark and antiquark are not too far apart, one can expect amplitudes of states with a large number of constituent gluons to be negligible, and can seek for the minimal energy eigenstate in a sector containing only the quark-antiquark pair plus a small number of constituent gluons.

So we will diagonalize $\mathfrak{T}$ in a finite $M$-dimensional subspace of trial ``chain'' states:
\begin{equation}\label{operators}
\vert \Psi_k\rangle=
\bar{q}^a(\mathbf{0})\;Q_k^{ab}[U]\;
q^b(\mathbf{R})\;
\vert\Psi_{0}\rangle
\qquad k=1,2,\dots,M,
\end{equation}
where $Q_k$ are gluonic operators, functionals of the lattice gauge field, that depend on quark/anti\-quark positions and some number of variational parameters (see the next section for a particular choice of the operator basis); $a$ and $b$ are color indices. All quantities of interest can be estimated from matrix elements
\begin{eqnarray}
O_{kl}= 
\langle \Psi_k
\vert\Psi_l\rangle&=&
\left\langle{\textstyle{\frac{1}{2}}}\mbox{Tr}
\left[Q^\dagger_k(t)Q_l(t)\right]\right\rangle,
\\
t_{kl}=
\langle \Psi_k\vert \mathfrak{T}
\vert\Psi_l\rangle&=&
\left\langle{\textstyle{\frac{1}{2}}}\mbox{Tr}
\left[Q^\dagger_k(t+1)U^\dagger_0(\mathbf{0},t)Q_l(t)U_0(\mathbf{R},t)\right]\right\rangle,
\end{eqnarray}
computed by lattice Monte Carlo simulations.\footnote{The notation $Q(t)$ indicates that the operator $Q[U]$ is evaluated using links on a hypersurface of fixed time $t$.} Knowing these matrix elements, one can construct, via the usual Gram--Schmidt procedure, 
an orthonormal set of states $\left\{\vert\Phi_k\rangle, k=1,2,\dots,M\right\}$,
then the matrix elements
\begin{equation}
T_{kl}=
\langle \Phi_k\vert \mathfrak{T}
\vert\Phi_l\rangle,
\end{equation}
and finally determine the largest eigenvalue $\tau_{\mathrm{max}}$ of the $M\times M$ matrix $T$. 
Such a calculation has to be repeated for various variational-parameter sets to determine the one which minimizes $-\log(\tau_{\mathrm{max}})$ at a given $R$.
An estimate of the static potential in the sector of variational gluon-chain states, nicknamed below the ``gluon-chain potential'', will then be given by
\begin{equation}\label{chain-potential}
V_\mathrm{chain}(R)=-\log(\tau_{\mathrm{max}}).
\end{equation}
%

\section{Choice of the operator basis}\label{basis}
Our choice of the operator basis in Eq.\ (\ref{operators}) was dictated mainly by simplicity, and by some amount of trial and error. 

A gluon chain is assumed to consist of a certain number of constituent gluons between the heavy quark and antiquark, and the gluon ordering in color indices is correlated with their spatial positions between the heavy color sources. As usual, in a variational approach the optimal energy states represent a compromise between kinetic energy and interaction energy. While the kinetic-energy contribution prefers spatial delocalization of gluons, the Coulombic interaction energy favors as small as possible transverse displacement from the line connecting the quark and antiquark positions. To satisfy both requirements, the delocalization in the spatial direction was achieved by a superposition of gluon operators in $Q_k$ along the line joining the sources, while delocalization in transverse directions could be realized by using ``transverse-smoothed'' gauge-field operators on the lattice, in which high-frequency components of the field are (e.g.) Gaussian-suppressed in the directions transverse to the $q\bar{q}$-line. The transverse smoothing introduced a single parameter $\rho$, the only variational parameter used in our operator Ansatz (see below). 
To further simplify this pilot study, we restricted the number of constituent gluons to at most two.

Our procedure was thus the following (for further details see Ref.\ \cite{Greensite:2009mi}):
\begin{itemize}
\item
The lattice configurations were fixed to the Coulomb gauge by standard methods. From lattice link matrices $U$ we constructed
\begin{equation}
A_i(\mathbf{x},t)={\textstyle{\frac{1}{2i}}}
\left[U_i(\mathbf{x},t)-U_i^\dagger(\mathbf{x},t)\right],\qquad
B_i(\mathbf{x},t)=1-{\textstyle{\frac{1}{2}}}
\mbox{Tr}\left[U_i(\mathbf{x},t)\right].
\end{equation}
\item
These quantities were then Fourier-transformed, and we suppressed high-momentum components in directions transverse 
to the line joining the $q\bar{q}$ pair (e.g.\ the $j$-th direction):
\begin{equation}
A_i(\mathbf{k},t)\to\exp\left[-\rho(\mathbf{k}^2-k_j^2)\right]
\;A_i(\mathbf{k},t),\qquad
B_i(\mathbf{k},t)\to\exp\left[-\rho(\mathbf{k}^2-k_j^2)\right]
\;B_i(\mathbf{k},t),
\end{equation}
then transformed them  back to coordinate space, to get $A_i^{(j)}(\mathbf{x},t)$ and $B_i^{(j)}(\mathbf{x},t)$, the $A$- and $B$-fields smeared in directions transverse to $\mathbf{e}_j$.
{$\rho$ is a variational parameter}, used to maximize the largest eigenvalue of the transfer matrix in a chosen  basis of states.
\item
It was also useful to define, for $i\ne j$, the averages:
\begin{eqnarray}
\bar{A}_i^{(j)}(\mathbf{x},t)&=&{\textstyle\frac{1}{2}}\left[A_i^{(j)}(\mathbf{x},t)+A_i^{(j)}(\mathbf{x}-\mathbf{e}_i,t)\right],\\
\bar{B}_i^{(j)}(\mathbf{x},t)&=&{\textstyle\frac{1}{2}}\left[B_i^{(j)}(\mathbf{x},t)+B_i^{(j)}(\mathbf{x}-\mathbf{e}_i,t)\right].
\end{eqnarray}
\item
Finally, a six-state basis was constructed from ``transverse-smoothed'' $A$- and $B$-fields that consisted of
\begin{eqnarray}
\label{onegluon}
\mbox{{\textbf{zero-gluon state}}: }&\dots&\
Q_1(t) = \mathbf{1}_2,\\
\mbox{{\textbf{one-gluon state}}: }&\dots&\
\label{twogluon}
Q_2(t) = \sum_{n=0}^{R-1} A_j^{(j)}(\mathbf{x}_0+n \mathbf{e}_j,t),\\  
\mbox{{\textbf{two-gluon states}}: }&\dots&\
Q_3(t) = \sum_{n=-2}^{R+1} ~ \sum_{n'=n}^{R+1} A_j^{(j)}(\mathbf{x}_0+n \mathbf{e}_j,t) A_j^{(j)}(\mathbf{x}_0+n' \mathbf{e}_j,t),
\nonumber \\
&\dots&\ 
Q_4(t) = \sum_{n=-2}^{R+2} ~ \sum_{n'=n}^{R+2} ~ \sum_{i\ne j}
         \overline{A}_i^{(j)}(\mathbf{x}_0+n \mathbf{e}_j,t) \overline{A}_i^{(j)}(\mathbf{x}_0+n' \mathbf{e}_j,t),
\nonumber \\
&\dots&\ 
Q_5(t) = \sum_{n=0}^{R-1} B_j^{(j)}(\mathbf{x}_0+n \mathbf{e}_j,t)\;\mathbf{1}_2,
\nonumber \\
\label{threegluon}
&\dots&\ 
Q_6(t) = \sum_{n=0}^{R-1} \sum_{i\ne j} \overline{B}_i^{(j)}(\mathbf{x}_0+n \mathbf{e}_j,t)\;\mathbf{1}_2.
\end{eqnarray}
The antiquark is assumed to sit at $\mathbf{x}_0$, the quark at 
$\mathbf{x}_0+R{\mathbf{e}}_j$. $Q_1$ is the zero-gluon operator, $Q_2$ the simplest one-gluon operator, with one $A$-field put at different locations between the quark and antiquark, and $Q_{3-6}$ are simple two-gluon operators, containing two powers of the gauge field $A$. (In two-gluon operators $Q_3$ and $Q_4$ the interval of $A$-field insertions was extended to up to two lattice spacings outside the region defined by quark/anti\-quark positions.)
\end{itemize}
Of course, one could use a larger set of more sophisticated operators and/or more variational parameters, but we believe that the above choice allowed to fulfill the goals of the study in a clear-cut and convincing, even though only qualitative, way.

\section{Results}\label{results}
The calculations outlined in Section \ref{outline} with the operators $Q_k$ given by Eqs.\ (\ref{onegluon}--\ref{threegluon}), were carried out for SU(2) lattice gauge theory
at coupling $\beta=2.2$ on a $12^4$ lattice volume, $\beta=2.3$ on a $16^4$ lattice volume, and $\beta=2.4$ on a $22^4$ lattice volume.  
The operators $Q_k$ depend implicitly on a variational parameter $\rho$, and matrix elements $T_{kl}$ and the gluon-chain potential
$V_\mathrm{chain}(R)$ were computed for each $R$ at twelve values of $\rho$, $\rho_n = (n-1) \Delta \rho$, $1\le n \le 12$,
with $\Delta \rho = 0.025$ at $\beta=2.2$, and $\Delta \rho=0.02$ at $\beta=2.3,2.4$.  The choice of $n$ which minimizes $V_\mathrm{chain}(R)$
depends on both $\beta$ and the quark separation $R$. For example, at $\beta=2.4$ and $R=9$, the optimal value was $n=8$.  
All our data were always obtained from the optimal value of $\rho$ for a given coupling and separation;
below we will, with the exception of Fig.\ \ref{fraction}, only report results for the largest coupling studied, $\beta=2.4$.

\subsection{Gluon-chain potential}
\begin{figure}[t!]
\centerline{\includegraphics[width=0.7\textwidth]{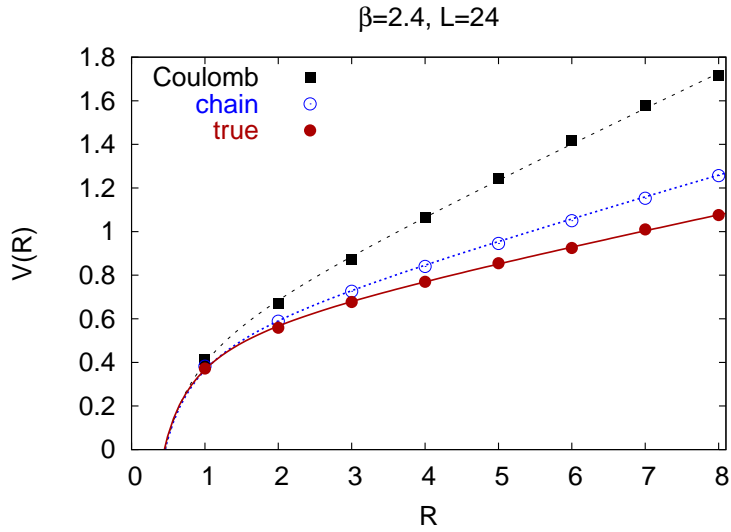}}
\caption{The color-Coulomb, gluon-chain, and ``true'' static quark-antiquark potentials vs.\ $R$ at $\beta=2.4$ on $22^4$ lattice.}\label{potentials}
\end{figure}

In Fig.\ \ref{potentials} we compare the color-Coulomb potential with the gluon-chain potential. The latter was estimated using 
Eq.\ (\ref{chain-potential}), while the former is given by $V_{\mathrm{Coul}}=-\log(T_{11})$, where $T_{11}$ is the zero-gluon to zero-gluon matrix element, independent of the variational parameter $\rho$. We also display the usual static quark-antiquark potential $V_{\mathrm{true}}$ computed by standard methods from timelike Wilson loops with ``fat'' spacelike links. The data were fitted by the usual (constant + L\"uscher + linear term) function, the extracted string tensions were 0.158, 0.095, 0.069 for the color-Coulomb, gluon-chain, and true potentials, respectively. The inclusion of one- and two-gluon operators affects the string tension in the expected way: the Coulomb string tension, about 2.3 times larger than the true asymptotic string tension (at $\beta=2.4$), goes down to the ``chain'' tension that differs from $\sigma_{\mathrm{asymp}}$ by 38\% only. One can imagine that a modest improvement of our operator basis plus inclusion of a few more constituent gluons would bring the string tension of the optimal variational gluon-chain state even closer to the true value. 

\subsection{Effects of finite volume}
\begin{figure}[b!]
\centerline{\includegraphics[width=0.7\textwidth]{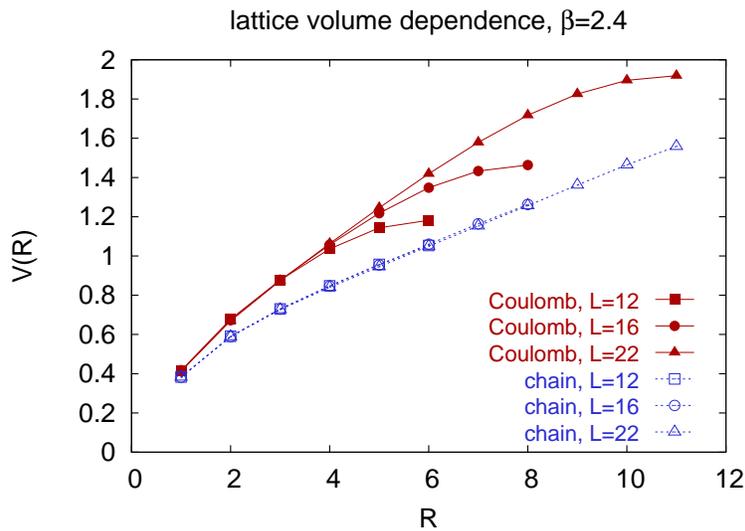}}
\caption{The color-Coulomb and chain potentials at different lattice volumes.}\label{finite-size}
\end{figure}
Figure \ref{finite-size} shows the dependence of the color-Coulomb potential and the gluon-chain potential on $R$ for three different lattice volumes. While $V_{\mathrm{Coul}}$ bends away from linearity at $R\approx L/2$, which is clearly a finite-size effect, $V_{\mathrm{chain}}$ seems completely insensitive to the size of the lattice. A natural interpretation of this result is that the long-range field does not exist or is greatly suppressed in the chain state relative to the color-dipole field of the zero-gluon state.

\subsection{Constituent gluon content of the gluon chain}
\begin{figure}[t!]
\centerline{\includegraphics[width=0.7\textwidth]{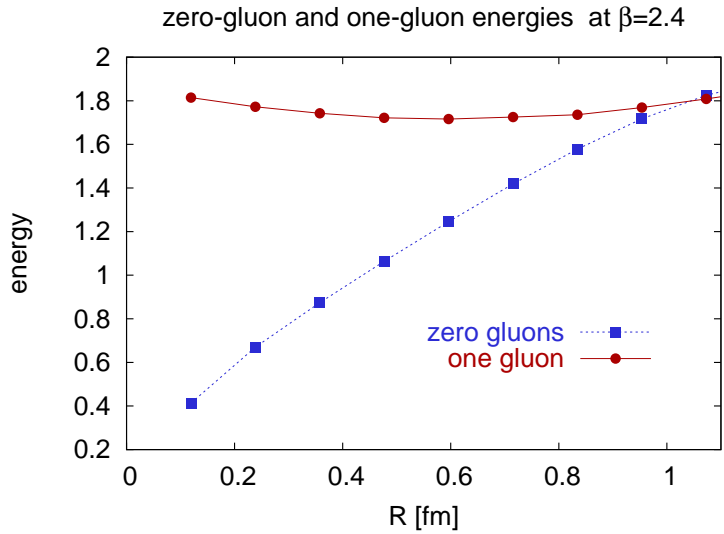}}
\caption{Energy of the zero-gluon and one-gluon states vs. quark-antiquark separation.}\label{energy}
\end{figure}
\begin{figure}[b!]
\centerline{\includegraphics[width=0.7\textwidth]{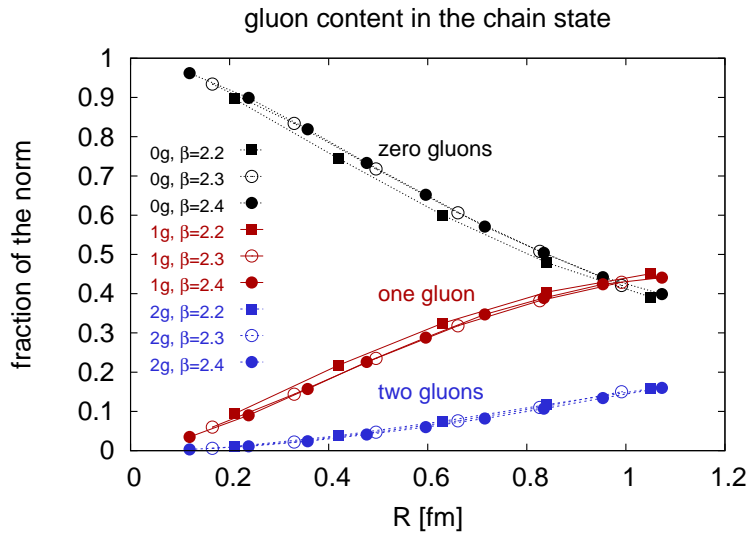}}
\caption{Zero-, one-, and two-gluon fraction of the norm of the
variational state vs.\ quark-antiquark distance $R$ at $\beta=2.2, 2.3, 2.4$.}\label{fraction}
\end{figure}

Finally we studied the constituent gluon content of our optimal variational gluon-chain state for different quark-antiquark separations $R$. In Figure \ref{energy} we display the $R$-dependence of the of the zero-gluon and one-gluon states at $\beta=2.4$. The energies are given by $-\log(T_{11})$, and $-\log(T_{22})$ respectively. The Coulombic energy of the zero-gluon state, and the kinetic plus interaction energy of the one-gluon state become 
equal at $R$ about 1 fm.

One can estimate the gluon content also directly. If we write the minimal-energy variational state in our six-state basis as
$\vert\Psi(R)\rangle=\sum_{k=1}^6 \alpha_k(R)\vert\Phi_k\rangle$, then the zero-, one-, and two-gluon fractions are given by 
$\alpha_1^2$, $\alpha^2_2$, and $1-\alpha_1^2-\alpha_2^2$, respectively. The results are shown in Figure \ref{fraction}.
The gluon content vs.\ $R$ (expressed in physical units) turns out to be almost independent of coupling.
The one-gluon content of the minimal energy state becomes equal to the zero-gluon content at about 1 fm, i.e.\ at the same distance at which the energies of zero- and one-gluon states equalize. 

\section{Conclusions}\label{conclusions}
A simple variational calculation of a quark-antiquark state in a subspace of zero, one, and two constituent-gluon states yields its energy less than 
the energy computed from a zero-gluon state. This result is not surprising by itself -- what appears nontrivial and was not guaranteed from the beginning is the following:

\begin{enumerate}
\item
The Coulombic energy of the zero-gluon state rises linearly with
separation, 
albeit with string tension higher than the asymptotic tension of the QCD flux tube.

\item
The linearity of the potential survives addition of a small number of constituent gluons.

\item
A few constituent gluons tend to bring the string tension of the variational state down considerably, to a value closer to the asymptotic string tension of the QCD flux tube.

\item
One begins to see the formation of the gluon chain only at quark-antiquark separations of about 1 fermi. 

\end{enumerate}


\end{document}